\documentclass[preprint,12pt]{elsarticle}

\usepackage{amssymb}
\usepackage{amsmath}

\journal{Computers in Biology and Medicine}

\usepackage{microtype} 
\usepackage{longtable}

\makeatletter
\def\ps@pprintTitle{%
 \let\@oddhead\@empty
 \let\@evenhead\@empty
 \let\@oddfoot\@empty
 \let\@evenfoot\@empty}
\makeatother

\begin{document}

\begin{graphicalabstract}
\includegraphics[width=1.0\textwidth]{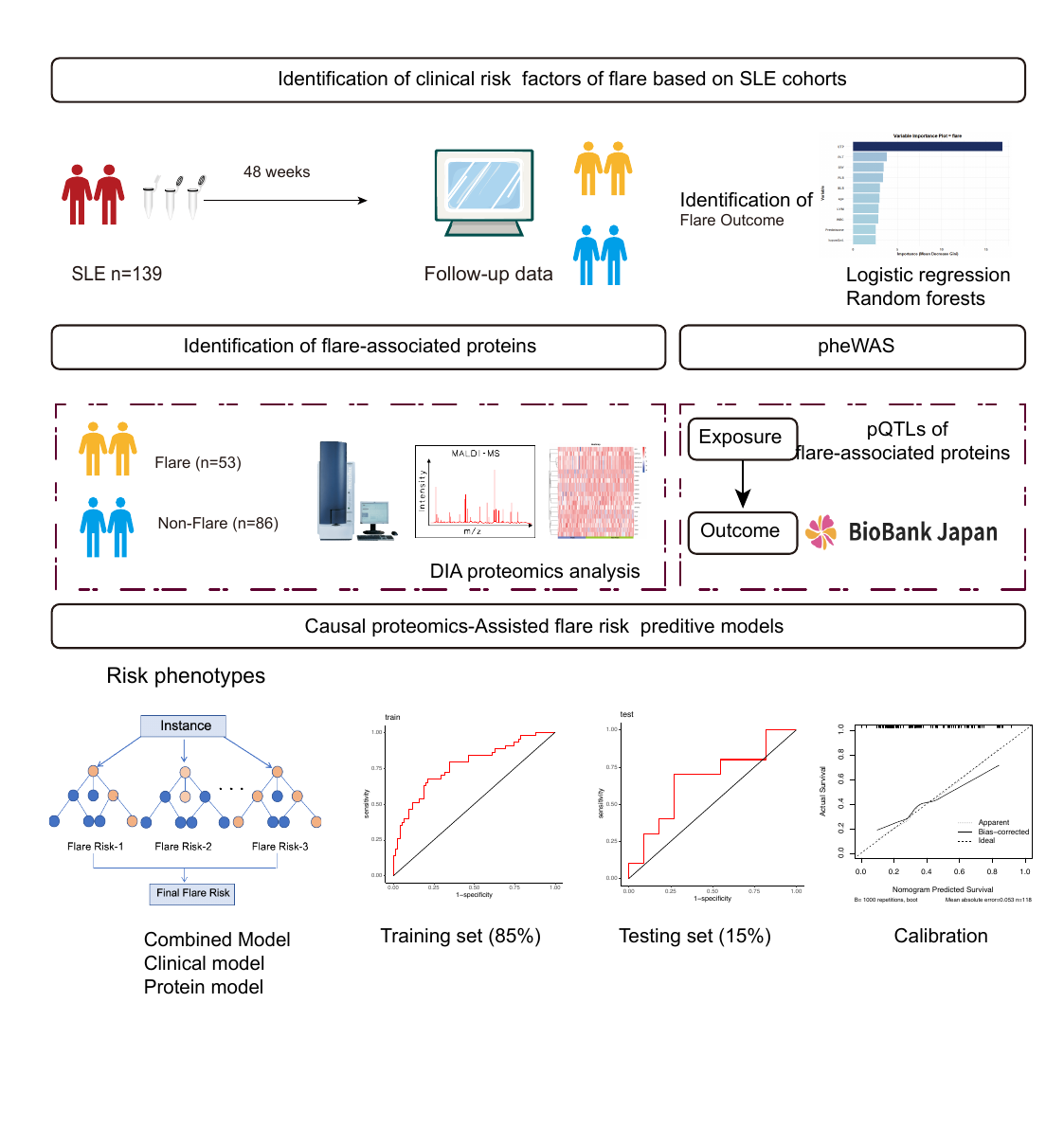}
\end{graphicalabstract}


\begin{frontmatter}



\title{Phenome-wide causal proteomics enhance systemic lupus erythematosus flare prediction: A study in Asian populations}

 \author[label1]{Liying Chen}
 \author[label3]{Ou Deng}
 
\author[label1]{Ting Fang} 
 \author[label1]{Mei Chen}
 \author[label1]{Xvfeng Zhang}
 
\author[label3]{Ruichen Cong} 
\author[label2]{Dingqi Lu} 
\author[label2]{Runrun Zhang}
\author[label3]{Qun Jin} 
\author{Xinchang Wang\corref{cor1}\fnref{label2}}
 
 \affiliation[label1]{organization={Second Clinical Medical College, Zhejiang Chinese Medical University},
             city={Hangzhou},
             postcode={310053},
             country={China}}
 \affiliation[label2]{organization={Department of Rheumatology, The Second Affiliated Hospital of Zhejiang Chinese Medical University},
             city={Hangzhou},
             postcode={310053},
             country={China}}

 \affiliation[label3]{organization={Graduate School of Human Sciences, Waseda University},
             addressline={Mikajima},
             city={Tokorozawa},
             postcode={359-1192},
             country={Japan}}



\begin{abstract}

\noindent \textbf{Objective}: Systemic lupus erythematosus (SLE) is a complex autoimmune disease characterized by unpredictable flares. This study aimed to develop a novel proteomics-based risk prediction model specifically for Asian SLE populations to enhance personalized disease management and early intervention.

\noindent \textbf{Methods}: A longitudinal cohort study was conducted over 48 weeks, including 139 SLE patients monitored every 12 weeks. Patients were classified into flare (n = 53) and non-flare (n = 86) groups. Baseline plasma samples underwent data-independent acquisition (DIA) proteomics analysis, and phenome-wide Mendelian randomization (PheWAS) was performed to evaluate causal relationships between proteins and clinical predictors. Logistic regression (LR) and random forest (RF) models were used to integrate proteomic and clinical data for flare risk prediction.

\renewcommand{\thefootnote}{}\footnotetext{\noindent Corresponding author: Xinchang Wang, Department of Rheumatology, The Second Affiliated Hospital of Zhejiang Chinese Medical University, No.318, Chaowang Road, Gongshu District, Hangzhou, Zhejiang Province, China. Email: 20054007@zcmu.edu.cn; ossani@126.com}

\noindent \textbf{Results}: Five proteins (SAA1, B4GALT5, GIT2, NAA15, and RPIA) were significantly associated with SLE Disease Activity Index-2K (SLEDAI-2K) scores and 1-year flare risk, implicating key pathways such as B-cell receptor signaling and platelet degranulation. SAA1 demonstrated causal effects on flare-related clinical markers, including hemoglobin and red blood cell counts. A combined model integrating clinical and proteomic data achieved the highest predictive accuracy (AUC = 0.769), surpassing individual models. SAA1 was highlighted as a priority biomarker for rapid flare discrimination.

\noindent \textbf{Conclusion}: The integration of proteomic and clinical data significantly improves flare prediction in Asian SLE patients. The identification of key proteins and their causal relationships with flare-related clinical markers provides valuable insights for proactive SLE management and personalized therapeutic approaches.

\end{abstract}


\end{frontmatter}



\section*{Introduction}
\label{sec1}

\noindent Systemic lupus erythematosus (SLE) is a heterogeneous autoimmune disorder characterized by the production of multiple autoantibodies and involvement of various organ systems. Its clinical course is unpredictable, with patients experiencing remissions and flares sudden increases in disease activity that pose significant management challenges\cite{1,2}. Although early disease control and adherence to pharmacological interventions are foundational in managing SLE, conventional predictive models relying on clinical observations and genetic markers are limited in capturing the disease's dynamic and multifactorial nature\cite{3,4}.

The complexity of SLE flares, often arising unpredictably from multifaceted and poorly understood triggers, necessitates a nuanced, data-driven predictive approach. Accurate flare prediction requires longitudinal tracking of disease activity, identification of modifiable risk factors, and comprehensive evaluation of clinical, serological, and molecular markers. Recent studies in Asian SLE populations have linked risk factors such as thrombocytopenia, hypocomplementemia, elevated neutrophil-to-lymphocyte ratio (NLR), and high platelet-to-lymphocyte ratio (PLR) to increased flare risk\cite{5,6}. Additionally, the presence of specific autoantibodies like anti-ribosomal P and anti-phospholipid antibodies is associated with heightened flare susceptibility in Chinese patients\cite{7}. Non-biological factors, including quality of life, psychological stress, and overexertion, also significantly contribute to flare incidence, highlighting the need to consider both biological and psychosocial variables in predictive models\cite{8,9,10}.

Advancements in genome-wide association studies (GWAS) and PheWAS have deepened our understanding of the genetic underpinnings and phenotypic expressions of autoimmune diseases. Mendelian randomization (MR) and PheWAS methodologies enable exploration of causal relationships between genetic variations, proteins, and disease outcomes, offering more robust insights than traditional observational studies\cite{11,12}. Given SLE's diverse nature, a multi-omics approach that integrates proteomic data with genomic and clinical information is essential for identifying dynamic biomarkers capable of improving flare prediction\cite{13}. In this context, proteomics-based MR holds significant promise for discovering novel, druggable protein targets, thereby enhancing our ability to classify and stratify patients by flare risk.

This study aims to address the limitations of current predictive models for SLE flares through a comprehensive, integrative approach. We first conduct an exhaustive analysis of known clinical predictors to build a robust baseline model. Building on this foundation, we incorporate cutting-edge plasma proteomics techniques to gain deeper insights into molecular changes associated with flare events. We explore causal relationships between identified plasma proteins and clinical outcomes through phenome-wide Mendelian randomization analyses, identifying proteins with a causal impact on flare risk. Finally, we leverage advanced machine learning algorithms to integrate proteomic and clinical data, constructing a predictive model that offers more precise risk stratification for SLE flares. This integrative methodology aims to significantly improve early detection of organ damage, guiding timely and personalized therapeutic interventions. 

\section*{Methods}
\label{sec2}

\subsection*{Patient Recruitment and Follow-up}
\noindent This prospective multicenter study recruited 139 SLE patients between August 2020 and January 2023. Eligible participants were aged 18-65 and met the 2012 SLICC classification criteria\cite{14}. Patients with active infections, malignancies, or other connective tissue diseases were excluded. Disease activity was assessed using the SLEDAI-2K at baseline and at 3, 6, and 12 months to identify flare events.

The study was approved by the ethics committee of the Second Affiliated Hospital of Zhejiang Chinese Medical University (approval NO.2020-KL-002-IH01). Informed consent was obtained from all participants. Additional details are in Supplementary Table 1.

\subsection*{Outcome Measurement}
\noindent  The primary outcome was the occurrence of an SLE flare within 12 months, defined as an increase in SLEDAI-2K score by $\geq 3$ points from baseline and previous assessments\cite{15}. This widely accepted criterion standardizes flare identification.

\subsection*{Clinical Variables: Definitions and Selection}
\noindent  Baseline data included: (1) Complement levels (C3, C4); (2) Inflammatory markers (NLR, PLR); (3) Medication use (glucocorticoids, immunosuppressants); (4) Serological markers (anti-ribosomal P antibodies); (5) Disease activity (SLEDAI-2K score); and (6) Quality of life (LupusQoL questionnaire). LR and RF algorithms identified key clinical predictors by ranking features based on predictive importance. All variables were dichotomized (1 or 0) for simplicity and clinical relevance, facilitating risk stratification. This approach aims to develop a robust predictive model integrating clinical parameters with proteomic biomarkers.

\subsection*{Proteomic Analysis}
\noindent  DIA data were processed into spectral libraries using SpectraST and analyzed with DIA-NN (v1.7.0), ensuring accurate results. Detailed workflows are available at https://www.iprox.cn/page/HMV006.html.

\subsection*{Differential Protein and SLEDAI-2K Correlation Analysis}
\noindent Proteins expressed in at least 25\% of samples were analyzed. Student's t-test identified plasma proteins differentiating flare and non-flare patients, considering proteins with $|\log_2(\text{Fold Change})| \geq 1$ and $p < 0.05$ as significant. Spearman's correlation assessed relationships between these proteins and SLEDAI-2K scores, with significance at $p < 0.05$.

\subsection*{Pathway Enrichment Analysis}
\noindent We performed pathway enrichment analysis on differentially expressed proteins and those correlated with SLEDAI-2K scores. We analyzed Reactome and KEGG pathways, considering those significantly enriched at $p < 0.05$ using DAVID (https://david.ncifcrf.gov/).

\subsection*{Causal Proteomics Analysis Using PheWAS}
\noindent Genetic instruments for identified pQTLs were obtained from 2,958 Han Chinese participants\cite{16}. Outcome data for SLE phenotypes were sourced from BioBank Japan ($n = 179{,}000$) and supplemented by UK Biobank and FinnGen data ($n_{\text{total}} = 628{,}000$)\cite{17}. The GWAS dataset included predictors like chronic glomerulonephritis, blood cell counts, and medication use. We selected pQTLs with $p < 5 \times 10^{-8}$, independence ($R^2 < 0.001$ with clumping distance $>10{,}000$ kb), and $F$ statistic $>10$ to ensure robust instruments.

The primary method was inverse-variance weighted Mendelian Randomization. To address potential violations, we performed sensitivity analyses using MR-Egger regression, weighted median, and mode-based estimators. Heterogeneity was assessed with Cochran's Q test, and leave-one-out analyses evaluated SNP influence on causal estimates. This approach allowed us to explore causal relationships between serum amyloid A1 (SAA1) pQTLs and SLE flare risk.

\subsection*{Machine Learning-Based Biomarker Selection}
\noindent We developed predictive models for SLE flares using LR and RF algorithms. Three models were created: (1) Clinical models using validated risk parameters; (2) Protein models using identified biomarkers; and (3) Combined models integrating clinical and proteomic data. Datasets were split into training (85\%) and testing (15\%) sets. Model performance was evaluated using Receiver Operating Characteristic curves and calibration curves to assess discriminative ability and reliability.

\subsection*{Statistical Analysis}
\noindent Statistical analyses were conducted using R (v4.3.2) and GraphPad Prism 8. Clinical data are presented as medians with interquartile ranges. Group comparisons used Student's t-test. Correlation analyses and linear regression assessed variable relationships. PheWAS analyses employed the \texttt{TwoSampleMR} R package. Predictive models used LR. Model accuracy was assessed using ROC curves and Area Under the Curve metrics via the \texttt{pROC} R package.

\section*{Results}

\subsection*{Clinical Data Characteristics and Flare Outcome}

\begin{figure}[h]
\begin{centering}
\includegraphics[width=0.9\linewidth]{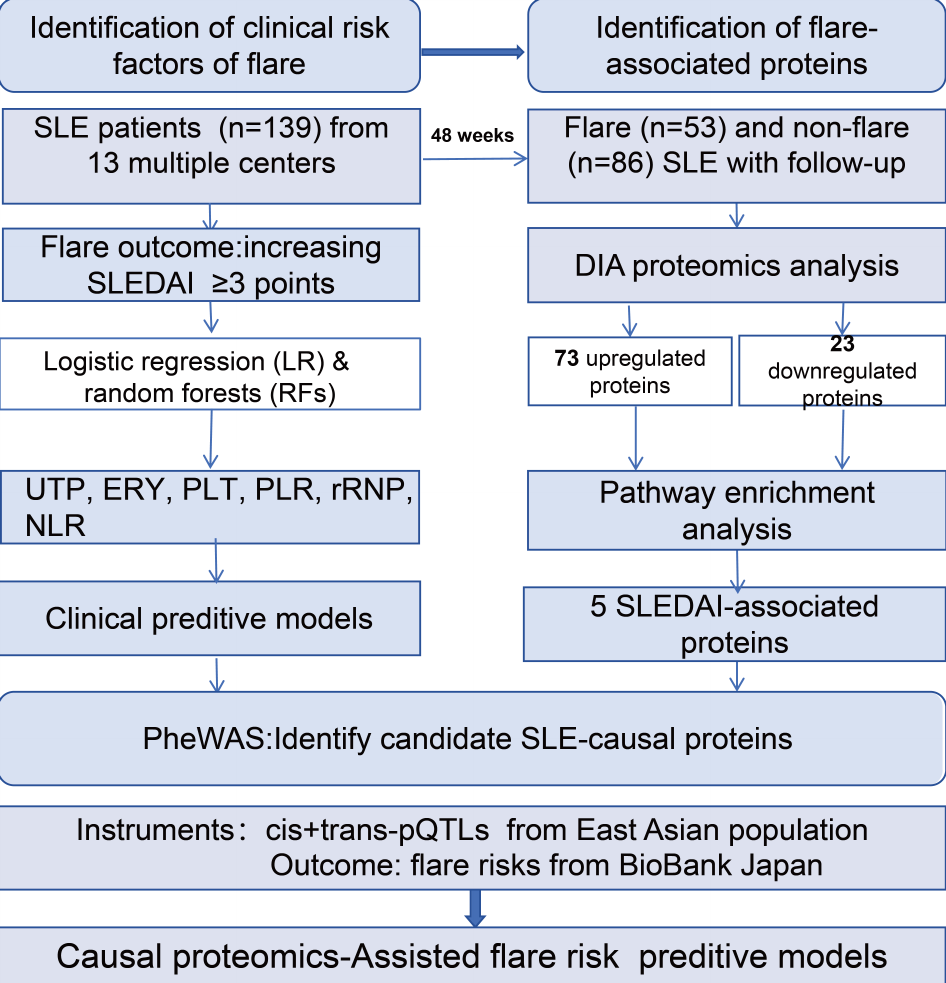}
\par\end{centering}
\caption{Study design and patient follow-up schema. Schematic representation of the recruitment process, inclusion and exclusion criteria, and longitudinal assessment timeline for SLE patients over the 48-week study period.\label{fig_1}}
\end{figure}

\noindent  As shown in Fig.\ref{fig_1}, 139 SLE patients were monitored over 48 weeks and classified into flare ($n=53$) and non-flare ($n=86$) groups; 38\% experienced at least one flare. The flare group had a high female predominance (94\%), aligning with SLE's gender disparity. Approximately 47\% of flare patients were positive for anti-dsDNA antibodies. Significantly, the flare group had a higher prevalence of rRNP antibodies than the non-flare group ($p=0.019$), suggesting their potential role in predicting flares.

Significant differences in erythrocyte sedimentation rate (ERY, $p=0.008$), platelet count (PLT, $p=0.023$), and platelet-to-lymphocyte ratio (PLR, $p=0.035$) were observed, highlighting their potential as flare risk markers. Medication use showed no significant difference in prednisone use ($p=0.284$). Hydroxychloroquine use was slightly lower in the flare group (by 11\%), though not statistically significant ($p=0.617$), warranting further investigation into its protective effects. No significant differences in organ involvement were observed ($p>0.05$). Baseline characteristics are detailed in Table\ref{tab_1}.

\begin{table}[h]
\centering
\caption{Clinical and demographic characteristics of SLE flare.\label{tab_1}}
\resizebox{\textwidth}{!}{%
\begin{tabular}{lcccc}
\hline 
Number of patients & Total ($n = 139$) & \multicolumn{3}{c}{Primary outcome: 1-year flare} \\
\cline{3-5}
 &  & Non-flare ($n = 86$) & Flare ($n = 53$) & $P$ \\
\hline 
Sex, n (\%) & 130 (94\%) & 80 (93\%) & 50 (94\%) & 1 \\
Age, Median (Q1, Q3) & 36 (28.5, 48.5) & 34 (31, 42.75) & 40 (28, 51) & 0.357 \\
Course, Median (Q1, Q3) & 6 (3, 12) & 6 (3, 12) & 6 (3, 11) & 0.617 \\
SLE\_DAI, Median (Q1, Q3) & 6 (3, 8) & 6 (3, 8) & 6 (4, 12) & 0.337 \\
dsDNA, n (\%) & 66 (47\%) & 41 (48\%) & 25 (47\%) & 1 \\
Sm, n (\%) & 35 (25\%) & 19 (22\%) & 16 (30\%) & 0.386 \\
SSA, n (\%) & 62 (45\%) & 37 (43\%) & 25 (47\%) & 0.763 \\
Ro52, n (\%) & 52 (37\%) & 28 (33\%) & 24 (45\%) & 0.185 \\
SSB, n (\%) & 22 (16\%) & 12 (14\%) & 10 (19\%) & 0.595 \\
rRNP, n (\%) & 38 (27\%) & 17 (20\%) & 21 (40\%) & 0.019 \\
Scl70, n (\%) & 1 (1\%) & 0 (0\%) & 1 (2\%) & 0.381 \\
AHA, n (\%) & 27 (19\%) & 17 (20\%) & 10 (19\%) & 1 \\
AnuA, n (\%) & 39 (28\%) & 24 (28\%) & 15 (28\%) & 1 \\
CENP, n (\%) & 2 (1\%) & 0 (0\%) & 2 (4\%) & 0.144 \\
AMA\_M2, n (\%) & 2 (1\%) & 1 (1\%) & 1 (2\%) & 1 \\
U10RNP, n (\%) & 25 (18\%) & 15 (17\%) & 10 (19\%) & 1 \\
pro, n (\%) & 48 (34\%) & 27 (31\%) & 21 (40\%) & 0.492 \\
ERY, Median (M25, M75) & 3 (0, 13) & 2.4 (0, 10.6) & 5.5 (1, 20) & 0.008 \\
LEU, Median (M25, M75) & 6 (2, 24.8) & 6 (1, 24.45) & 8 (2.3, 26) & 0.333 \\
UTP, Median (M25, M75) & 102.45 (102.45, 180) & 102.45 (102.45, 102.45) & 180 (180, 180) & $<$ 0.001 \\
WBC, Median (M25, M75) & 5.2 (4.3, 6.86) & 5.2 (4.27, 6.9) & 5.1 (4.4, 6.4) & 0.873 \\
NEUT, Median (M25, M75) & 3.43 (2.46, 4.82) & 3.46 (2.5, 4.78) & 3.33 (2.28, 4.88) & 0.714 \\
LYM, Median (M25, M75) & 1.36 (1.04, 1.97) & 1.33 (1.03, 1.88) & 1.43 (1.04, 2.23) & 0.534 \\
PLT, Median (M25, M75) & 208 (157, 248.5) & 217 (172.5, 258.75) & 175.5 (138, 225) & 0.023 \\
PLR, Median (M25, M75) & 145.37 (102.39, 190.67) & 156.1 (111.63, 200.63) & 134.18 (83.08, 179.71) & 0.035 \\
NLR, Median (M25, M75) & 2.36 (1.62, 3.51) & 2.37 (1.79, 3.48) & 2.25 (1.51, 3.54) & 0.411 \\
C3, Median (M25, M75) & 0.76 (0.66, 0.9) & 0.74 (0.66, 0.9) & 0.78 (0.67, 0.93) & 0.606 \\
C4, Median (M25, M75) & 0.15 (0.11, 0.2) & 0.15 (0.11, 0.2) & 0.15 (0.1, 0.2) & 0.776 \\
lupusQoL, Median (M25, M75) & 114 (101, 121) & 114 (103.25, 120.75) & 110 (97, 122) & 0.529 \\
Prednisone, Median (M25, M75) & 10 (5, 15) & 7.5 (5, 15) & 10 (5, 15) & 0.284 \\
Hydroxychloroquine, n (\%) & 86 (62\%) & 57 (66\%) & 47 (55\%) & 0.617 \\
Immunosuppressant, n (\%) & 90 (65\%) & 57 (66\%) & 33 (62\%) & 0.765 \\
Noninvolved, n (\%) & 31 (22\%) & 17 (20\%) & 14 (26\%) & 0.481 \\
Multisystem, n (\%) & 40 (29\%) & 27 (31\%) & 13 (25\%) & 0.499 \\
Hematologic, n (\%) & 22 (16\%) & 15 (17\%) & 7 (13\%) & 0.671 \\
Neuropsychiatric, n (\%) & 7 (5\%) & 3 (3\%) & 4 (8\%) & 0.427 \\
Mucocutaneous, n (\%) & 62 (45\%) & 36 (42\%) & 26 (49\%) & 0.514 \\
Renal, n (\%) & 57 (41\%) & 39 (45\%) & 18 (34\%) & 0.251 \\
\hline 
\end{tabular}
}
\end{table}

\subsection*{Importance Ranking of Clinical Predictors}

\begin{figure}[h]
\begin{centering}
\includegraphics[width=1.0\linewidth]{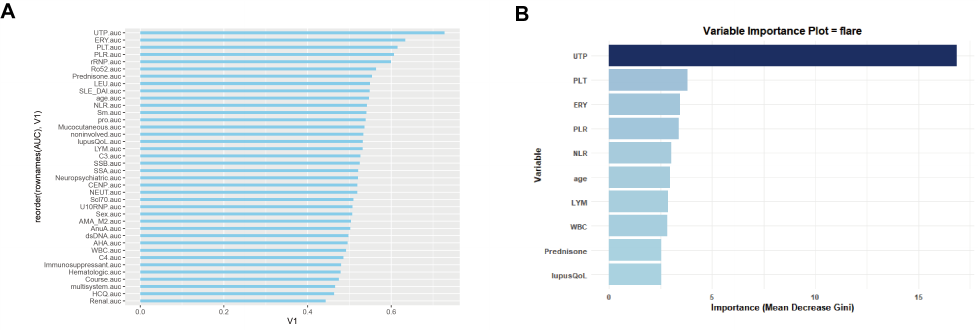}
\par\end{centering}
\caption{Comparative analysis of clinical variable importance in predicting SLE flares. (A) Ranking of clinical variables based on LR coefficients. (B) Relative importance of clinical variables determined by the RF algorithm, measured by the mean decrease in Gini impurity.\label{fig_2}}
\end{figure}

\noindent  We identified key clinical risk factors for SLE flares using univariate LR and RF based on 39 baseline features from 139 patients. In the LR model (Fig.\ref{fig_2}A), top predictors were: (1) 24-hour urinary total protein (UTP), (2) urinary erythrocyte count (ERY), (3) platelet count (PLT), (4) platelet-to-lymphocyte ratio (PLR), and (5) anti-ribonucleoprotein (rRNP) antibodies. The RF model (Fig.\ref{fig_2}B) yielded a similar ranking, with NLR as the fifth predictor.

\subsection*{Identification of Differentially Expressed Biomarkers Associated with Flare}

\begin{figure}[h]
\begin{centering}
\includegraphics[width=1.0\linewidth]{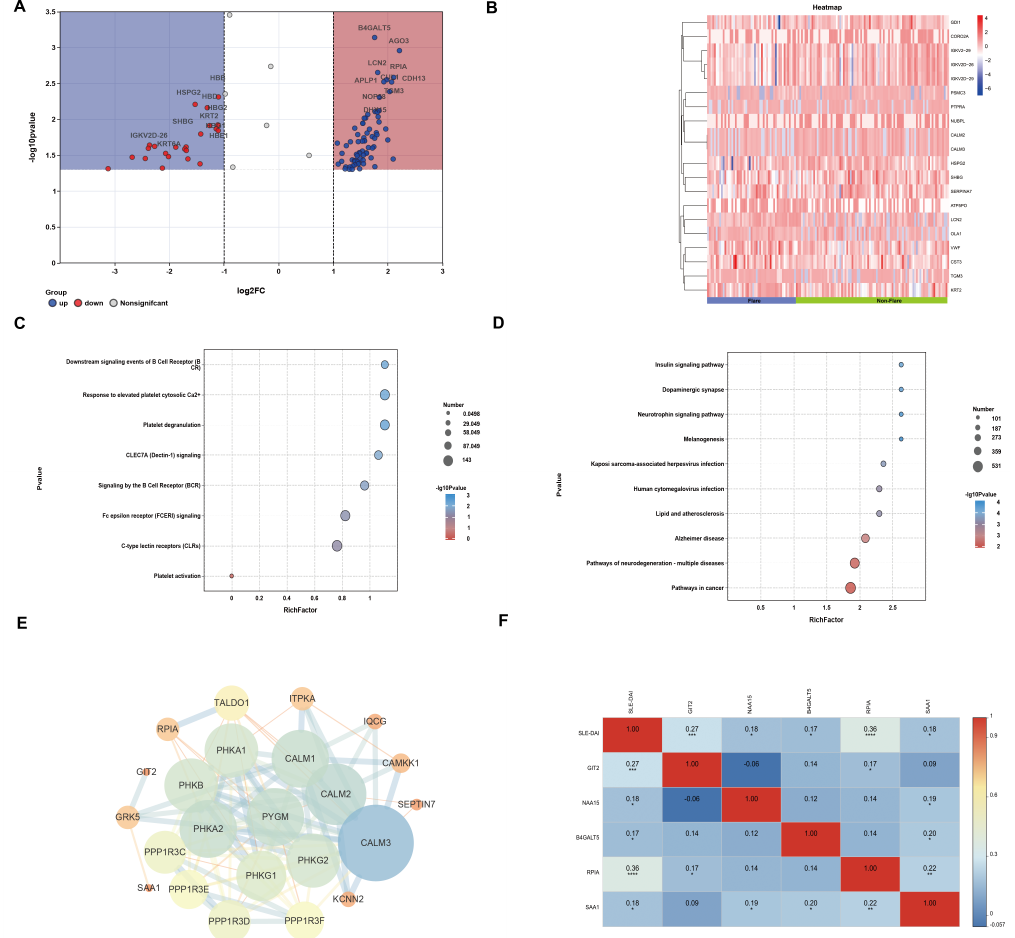}
\par\end{centering}
\caption{Multi-dimensional proteomic profiling of SLE flares. (A) Volcano plot illustrating the distribution and statistical significance of 102 differentially expressed proteins between flare and non-flare groups (`the absolute value of the base-2 logarithm of the fold change' $\ge$ 1,  $p < 0.05$). (B) Hierarchical clustering heatmap displaying normalized plasma protein expression levels, with a color gradient (blue to red) representing relative protein abundance. (C) Reactome pathway enrichment analysis of flare-associated proteins, with statistical significance indicated by -log10(p-value). (D) KEGG pathway analysis of proteins correlated with SLEDAI-2K scores. (E) STRING-based protein-protein interaction network of SLEDAI-2K-associated proteins, where node size reflects the degree of interaction. (F) Correlation matrix showing the relationships between upregulated proteins positively associated with SLEDAI-2K scores.\label{fig_3}}
\end{figure}

\noindent  Proteomic analysis identified 102 significantly differentially expressed proteins (73 upregulated, 23 downregulated) associated with flares (Fig.\ref{fig_3}A), based on proteins detected in over 25\% of samples. A heatmap (Fig.\ref{fig_3}B) illustrates these findings. Pathway enrichment analysis using Reactome revealed involvement in key pathways: (1) B-cell receptor (BCR) downstream signaling, (2) response to elevated platelet cytosolic $\mathrm{Ca}^{2+}$, and (3) platelet degranulation (Fig.\ref{fig_3}C), offering insights into molecular mechanisms underlying flares.

\subsection*{Proteins Significantly Associated with SLEDAI-2K}

\noindent  We identified plasma proteins significantly associated with SLEDAI-2K scores. KEGG pathway analysis indicated their involvement in neurodegeneration, lipid metabolism, atherosclerosis, and neurotrophin signaling (Fig.\ref{fig_3}D). Protein-protein interaction analysis using STRING (Fig.\ref{fig_3}E) showed functional relationships among these proteins. Among 13 proteins correlating with SLEDAI-2K, five were significantly upregulated and positively associated with worse outcomes: (1) Serum Amyloid A-1 (SAA1), (2) $\beta$-1,4-galactosyltransferase 5 (B4GALT5), (3) Ribose 5-phosphate Isomerase A (RPIA), (4) GTPase-activating Protein 2 (GIT2), and (5) N-alpha-acetyltransferase 15 (NAA15). Correlation analysis (Fig.\ref{fig_3}F) showed significant positive associations between SAA1 and B4GALT5, RPIA, and NAA15, suggesting potential functional relationships in SLE activity and flare risk.

\subsection*{Causal Effects of Flare-Associated Proteins on SLE and Risk Factors}

\begin{figure}[h]
\begin{centering}
\includegraphics[width=1.0\linewidth]{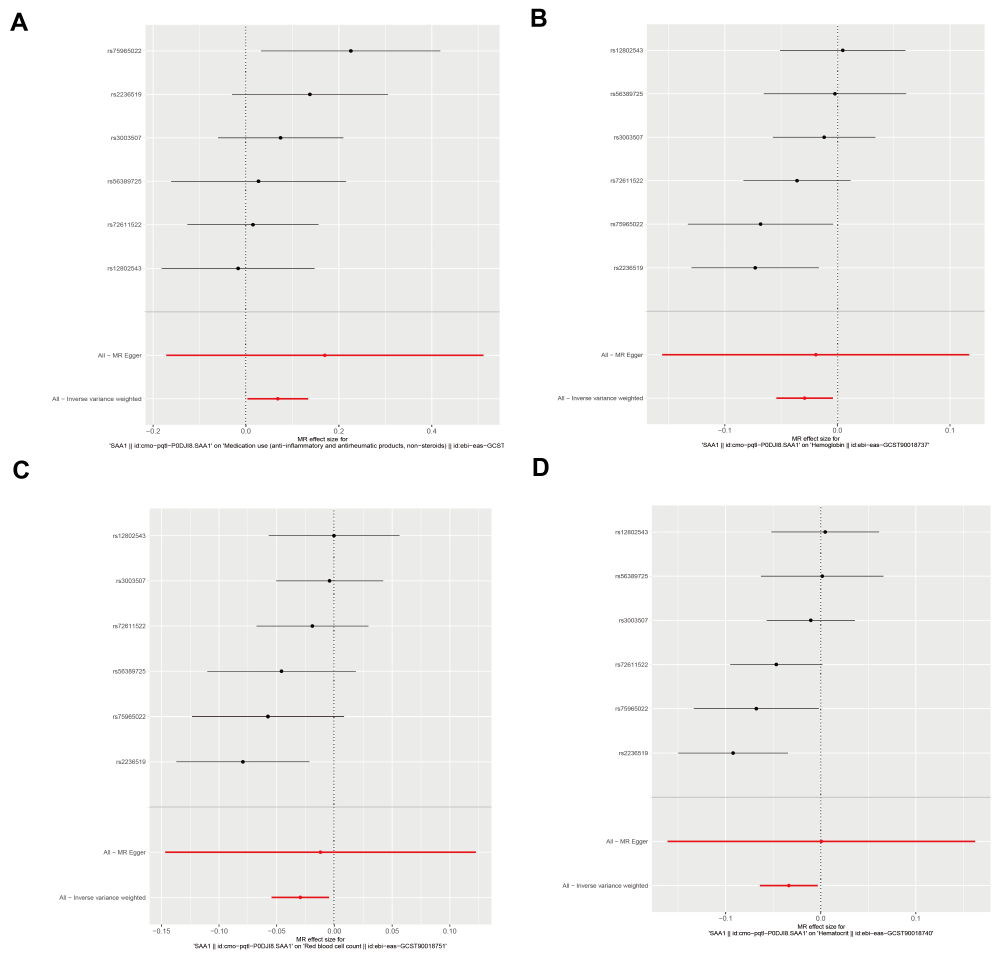}
\par\end{centering}
\caption{Mendelian randomization analysis of SAA1's causal effects on SLE flare risk factors. Forest plots from random-effects inverse-variance weighted (IVW) analyses depicting the genetic causal relationships between SAA1 and (A) anti-inflammatory and antirheumatic medication use, (B) hemoglobin levels, (C) red blood cell count, and (D) hematocrit. Odds ratios and 95\% confidence intervals are shown.\label{fig_4}}
\end{figure}

\begin{figure}[h]
\begin{centering}
\includegraphics[width=1.0\linewidth]{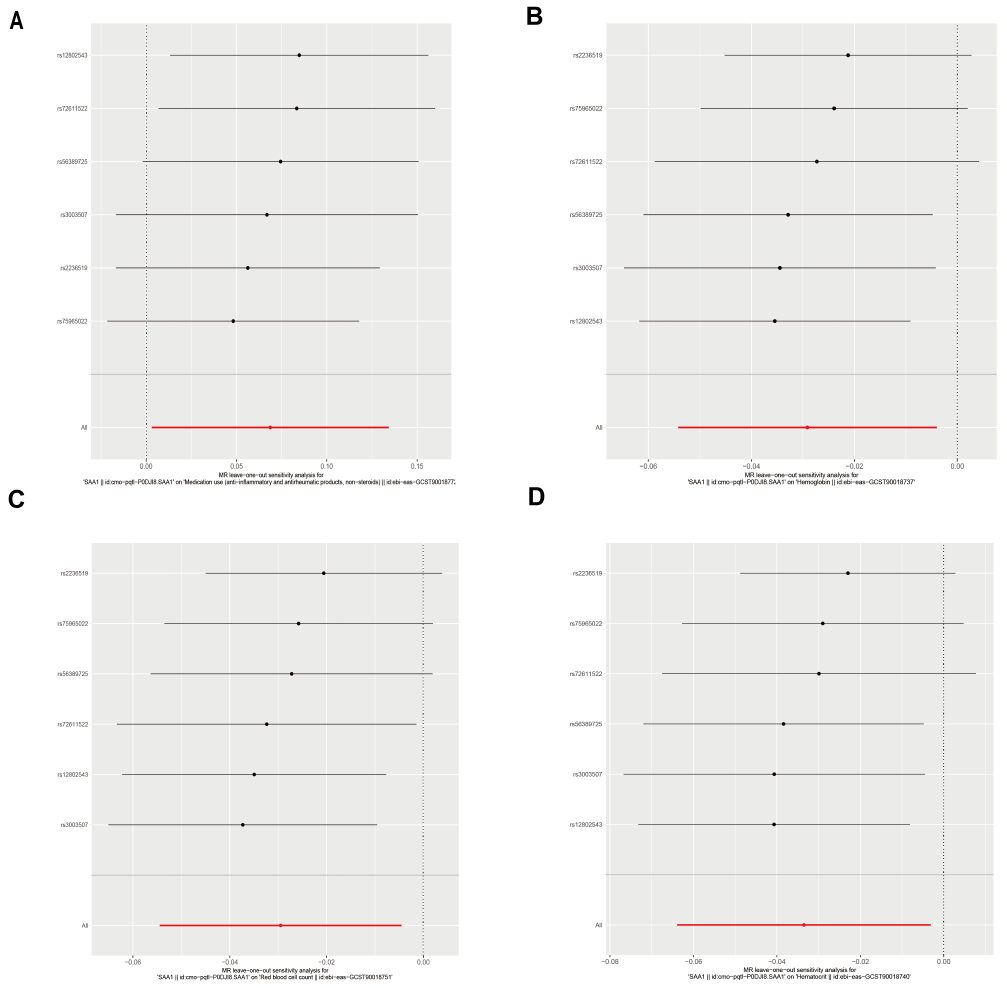}
\par\end{centering}
\caption{Sensitivity analysis of Mendelian randomization results using the leave-one-out method. (A) Anti-inflammatory and antirheumatic medication use. (B) Hemoglobin levels. (C) Red blood cell count. (D) Hematocrit. This analysis evaluates the robustness of the causal relationships between SAA1 and four key outcomes: anti-inflammatory and antirheumatic medication use, hemoglobin levels, red blood cell count, and hematocrit. Each subplot (A-D) represents one of these outcomes, with individual points showing the Mendelian randomization estimate after excluding one SNP from the analysis, while the vertical line indicates the estimate using all SNPs. This approach helps identify potential outlier SNPs that may disproportionately influence the overall results.\label{fig_5}}
\end{figure}

\noindent  Using PheWAS, we investigated causal mechanisms between plasma proteins and SLE flares, focusing on pQTLs for SAA1. We examined 220 SLE-related outcomes using data from BioBank Japan. Six SNPs associated with SAA1 levels were identified as instrumental variables. Inverse-variance weighted (IVW) analysis revealed significant causal effects of SAA1 on four flare-related outcomes (OR = 1.071, 95\% CI: 1.004 - 1.143, p = 0.040), hemoglobin (OR= 0.971, 95\% CI: 0.947 - 0.996, $p = 0.023$), red blood cell count (OR = 0.971, 95\% CI: 0.947 - 0.996, $p = 0.021$), hematocrit (OR =0.967, 95\% CI: 0.937 - 0.997, $p = 0.031$) (Fig.\ref{fig_4}A-D, Table \ref{tab_2}). All of them exhibited absence of heterogeneity by IVW (Cochran's Q = 0.419; Cochran's Q = 0.263; Cochran's Q = 0.291
; Cochran's Q =0.104, respectively) and MR-Egger (Cochran's Q= 0.334; Cochran's Q = 0.169; Cochran's Q = 0.195; Cochran's Q =0.068, respectively), horizontal pleiotropy by MR-Egger (intercept = -0.015, p= 0.585; intercept = -0.001, $p = 0.891$; intercept = -0.003, p = 0.808; intercept = -0.005, $p = 0.694$, respectively). The leave-one-out method suggested that the MR analysis results were reliable (Fig.\ref{fig_5}A-D).


\begin{table}[h]
\centering
\caption{Mendelian Randomization Results - Causal Effects Between pQTLs and SLE Flare Risk (IVW Approach)\label{tab_2}}
\resizebox{\textwidth}{!}{%
\begin{tabular}{llrrrr}
\hline
Outcome & N SNP & \multicolumn{1}{c}{$\beta$} & \multicolumn{1}{c}{Standard Error} & \multicolumn{1}{c}{$p$} & OR (95\% CI) \\
\hline
 Hemoglobin & 6 & -0.029 & 0.013 & 0.023 & 0.971 (0.947, 0.996) \\
 Hematocrit & 6 & -0.034 & 0.016 & 0.031 & 0.967 (0.937, 0.997) \\
 Red blood cell count & 6 & -0.029 & 0.013 & 0.021 & 0.971 (0.947, 0.996) \\
 Anti-inflammatory medication & 6 & 0.069 & 0.033 & 0.040 & 1.071 (1.004, 1.143) \\
\hline
\textit{Note: Source by Zheng Lab.}
\end{tabular}
}
\end{table}

\subsection*{Construction and Internal Validation of Flare Risk Prediction Models}

\noindent  We developed predictive models for SLE flare risk by dividing patients into training and test sets. Using LR and random forest algorithms, and following TRIPOD guidelines\cite{18}, we selected six key clinical variables: UTP, ERY, PLT, PLR, anti-rRNP antibodies, and NLR.

Three models were developed: a clinical model, a protein-based model using five key proteins (SAA1, B4GALT5, GIT2, NAA15, RPIA), and a combined model integrating clinical and proteomic variables. 
ROC curve analysis showed the protein-based model had superior predictive accuracy (AUC = 0.744, 95\% CI: 0.646 - 0.842) compared to the clinical model  (AUC = 0.643, 95\%CI: 0.541 - 0.745). The combined model achieved the highest AUC (AUC = 0.769, 95\% CI : 0.678 - 0.860), indicating the benefit of integrating clinical and proteomic data.

\begin{figure}[h]
\begin{centering}
\includegraphics[width=1.0\linewidth]{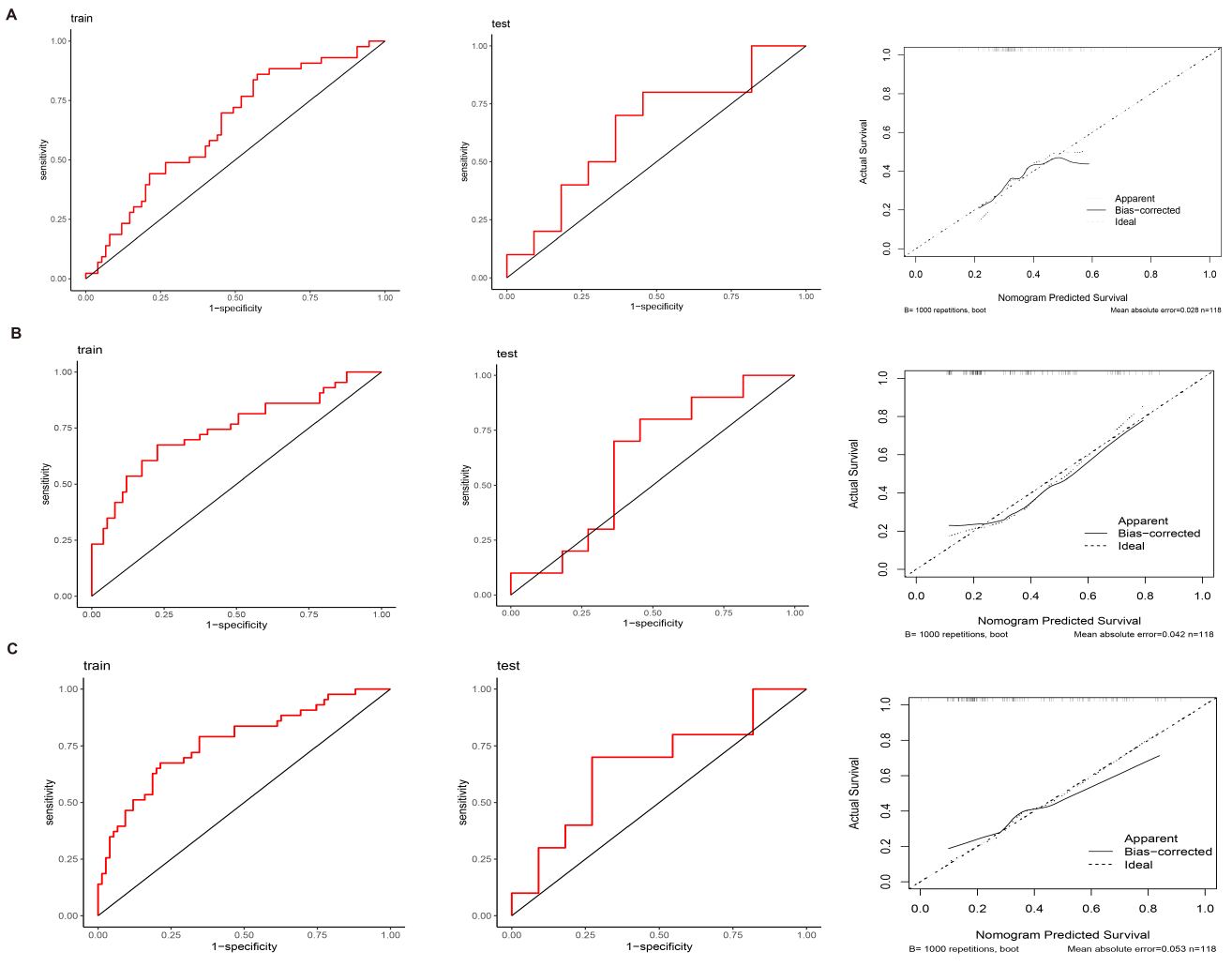}
\par\end{centering}
\caption{Comparative performance evaluation of SLE flare risk prediction models. (A-C) Receiver Operating Characteristic (ROC) curves (left) and calibration plots (right) for (A) clinical, (B) proteomic, and (C) combined prediction models. ROC curves display sensitivity versus 1-specificity, with Area Under the Curve (AUC) values indicated. Calibration plots show predicted versus observed probabilities, generated using 1,000 bootstrap resamples. Hosmer-Lemeshow goodness-of-fit test P-values are provided to evaluate model calibration.\label{fig_6}}
\end{figure}

Calibration analysis demonstrated good predictive performance for all models (Fig.\ref{fig_6}A-C). The protein-based model showed particular strength in AUC and Mean Absolute Error (MAE), suggesting its suitability for independent flare prediction. The combined model may be more appropriate in complex clinical scenarios requiring multiple variables for accurate risk assessment.

\section*{Discussion}

\noindent  This is the first study to develop an SLE flare prediction model based on proteomic analysis in an East Asian population. We identified novel molecular pathways and biomarkers associated with SLE flares, offering valuable insights for improved risk stratification and personalized management. 

Pathway enrichment analysis showed significant involvement of B cell receptor signaling, elevated platelet cytosolic $\mathrm{Ca}^{2+}$ responses, and platelet degranulation in SLE flare pathogenesis. KEGG analysis revealed enrichment of neurodegeneration, lipid metabolism, atherosclerosis, and neurotrophin signaling pathways among proteins linked to disease activity. These findings highlight the complex interplay between immune activation, vascular responses, and systemic inflammation in SLE flares, particularly affecting hematological, cardiovascular, and neurological systems in this East Asian cohort.

We identified five key proteins, SAA1, B4GALT5, GIT2, NAA15, and RPIA, strongly associated with increased SLE flare risk over 48 weeks, independent of conventional risk factors. Our analysis demonstrated a causal effect of elevated SAA1 levels on flare risk factors, underscoring its critical role in disease progression.

Integrating these proteomic biomarkers with clinical indicators significantly enhanced our model's predictive accuracy. These results confirm the prognostic potential of causal proteomics in SLE flare risk stratification, paving the way for personalized therapies and earlier interventions for high-risk patients.


Previous studies support Serum Amyloid A1 (SAA1) as a key biomarker for assessing SLE flare risk, reinforcing its predictive utility. SAA1-associated pQTLs have been causally linked to hematological parameters and depressive disorders in multi-GWAS PheWAS\cite{19}, enhancing our understanding of proteins in the psychological aspects of SLE flares\cite{20}. As an acute-phase protein, SAA1 is markedly upregulated during inflammation, contributing to organ dysfunction\cite{21}. Our study showed that elevated SAA1 levels causally influence hematological involvement and antirheumatic medication use, consistent with prior findings\cite{22}.

Elevated SAA1 levels correlate with SLEDAI-2K scores and nervous system involvement severity\cite{23}. SAA1 is linked to the Th17 cell differentiation pathway, amplifying inflammatory responses\cite{24}. Given its pivotal role in modulating inflammatory mediators and immune cells in SLE pathophysiology\cite{25}, investigating SAA1 can advance our understanding of flare mechanisms.

Identified as an early immunological diagnostic biomarker with high sensitivity and specificity for SLE\cite{26}, SAA1 may serve as a valuable biomarker for identifying high-risk patients based on our East Asian PheWAS results. These findings suggest potential therapeutic targets for personalized treatments. However, since infections also trigger flares and SAA1 is associated with white blood cell count and NLR, further research is needed to distinguish SLE flares from infection-related events\cite{27}.

$\beta$-1,4-Galactosyltransferase V (B4GALT5) is crucial in carbohydrate metabolism, specifically in lactosylceramide synthesis \cite{28}. Unexpectedly, our logistic regression showed a strong positive correlation between B4GALT5 expression and SLE flares, making it a significant protein associated with flare risk. Given B4GALT5's role in antiviral immunity, overexpression leads to upregulation of inflammatory cytokines and glycosylated surface proteins involved in antigen presentation, cell adhesion, and migration\cite{29}.

However, we did not establish a causal relationship between B4GALT5 and SLE, possibly due to the low prevalence of B4GALT5 variants in the East Asian population studied. This underscores the need for diverse population studies to evaluate B4GALT5's role in SLE pathogenesis and its predictive value.
The unexpected correlation suggests a complex interplay between glycosylation processes, immune responses, and SLE activity. Further investigation is needed to elucidate how B4GALT5 might influence flare risk, given its functions in carbohydrate metabolism and immune regulation. These findings open avenues for understanding the glycosylation-immune axis in SLE and potential therapeutic interventions.


GIT2 and NAA15 emerged as potential contributors to SLE flares. GIT2 is a scaffold protein regulating aging processes affecting multiple tissues and linked to neurodegeneration and cardiovascular disorders\cite{30}. NAA15 is a susceptibility gene for neurodevelopmental disorders\cite{31}, implicated in congenital cardiac anomalies, plaque stability in atherosclerosis, and seizure pathophysiology\cite{32,33}. Although no direct evidence links GIT2 and NAA15 to SLE, their identification suggests novel pathways in SLE pathogenesis. Further research is needed to elucidate their specific roles in SLE flares, particularly considering disease duration and age, which merit attention in risk prediction models.

SLE progression involves distinct risk factors at different disease stages. Early organ damage is driven by active disease (reflected in SLEDAI-2K), while later damage results from long-term medication effects, especially prolonged glucocorticoid use and withdrawal\cite{34,35,36}. We identified key clinical risk factors for flares, including hematological and renal involvement and medication use, aligning with traditional findings. Risk factors may shift during disease progression, with early damage from disease activity and later damage from drug side effects, notably glucocorticoid therapy.

Traditional correlation studies offer insights but do not establish causality. We employed proteomic Mendelian randomization to develop robust flare prediction models. By integrating correlational methods with advanced machine learning and causal analyses, we aimed to enhance prediction accuracy. Identifying proteomic biomarkers with direct causal effects provides deeper insights into SLE flare mechanisms, highlighting potential causal pathways and therapeutic targets for future interventions.

This study has several limitations. The small sample size and focus on an East Asian population may limit statistical power and generalizability to broader populations. The 12-month follow-up may be insufficient to capture long-term fluctuations in SLE disease activity. Moreover, reliance on advanced proteomics techniques, while valuable, may hinder clinical accessibility and routine implementation. Although our PheWAS analysis revealed associations between genetic protein levels and SLE, establishing definitive causality remains challenging due to the complexities of SLE and potential violations of Mendelian randomization assumptions. Therefore, these results should be interpreted with caution.
Future research should prioritize external validation of our predictive model in diverse cohorts, adhering to the TRIPOD guidelines. Studies with larger, heterogeneous populations and extended follow-up periods are essential to address current limitations. Developing methodologies that account for time-varying exposures and non-linear associations will enhance model robustness and accuracy. Further investigation is needed to verify causal relationships between identified proteins and SLE risk, improving clinical applicability. Additionally, comprehensive proteomic analyses across multiple organs are necessary to assess organ-specific protein effects, providing deeper insights into early flare mechanisms.

\section*{Conclusion}

\noindent This study demonstrates the potential of integrating causal proteomics with clinical risk factors to improve SLE flare prediction. Our findings provide significant advancements in understanding the molecular mechanisms underlying SLE flares and offer a foundation for more precise and personalized approaches to SLE management. The identified proteomic signatures and causal pathways represent promising avenues for future therapeutic interventions and risk stratification strategies, with the potential to enhance patient care and outcomes in SLE.

\vspace{1em}

\noindent {\footnotesize 
\textbf{Abbreviations:} \\ 
SLE: systemic lupus erythematosus; \\
PheWAS: phenome-wide mendelian randomization; \\
GWAS: genome-wide association studies;\\
MR: mendelian randomization;\\
DIA: data-independent acquisition;\\
LR: logistic regression;\\
RF: random forest;\\
UTP: 24-hour urinary total protein;\\
ERY: urinary erythrocyte count;\\
PLT: platelet count;\\
NLR: neutrophil-to-lymphocyte ratio; \\
PLR: platelet-to-lymphocyte ratio;\\
rRNP: anti-ribosomal P antibodies;\\
LupusQoL: lupus quality of life;\\
SLEDAI-2K: Systemic Lupus Erythematosus Disease Activity Index 2000;\\
BCR: B-cell receptor; \\
SLICC: systemic lupus international collaborating clinics;\\
KEGG: Kyoto Encyclopedia of Genes and Genomes;\\
SAA1: serum amyloid A1; \\
B4GALT5: $\beta -1$,4-galactosyltransferase 5;\\
RPIA: Ribose 5-phosphate isomerase A;\\
GIT2: GTPase-activating protein 2;\\
NAA15: N-alpha-acetyltransferase 15;\\
IVW: inverse-variance weighted;\\
TRIPOD: transparent reporting of a multivariable prediction model for individual prognosis or diagnosis.
}

\clearpage 

\section*{}

\renewcommand{\appendixname}{Supplementary Tables}

\clearpage 

\appendix

\section*{Supplementary Tables}

\subsection*{1. Name of multi-center hospitals}

\begin{longtable}{|p{0.9\textwidth}|}
    \hline
    \textbf{Table S1. Name of multi-center hospitals} \\ \hline
    The Second Affiliated Hospital of Zhejiang Chinese Medical University \\ \hline
    Guang’anmen Hospital China Academy of Chinese Medical Sciences \\ \hline
    Chinese People's Liberation Army General Hospital \\ \hline
    The First Affiliated Hospital of Guangzhou University of Chinese Medicine \\ \hline
    Shenzhen Futian Hospital for Rheumatic Diseases \\ \hline
    The First Affiliated Hospital of Henan University of CM \\ \hline
    First Teaching Hospital of Tianjin University of Traditional Chinese Medicine \\ \hline
    The First School of Clinical Medicine Yunnan University of Chinese Medicine \\ \hline
    Department of Rheumatology Second Affiliated Hospital of Zhejiang University \\ \hline
    The Affiliated Hospital of Medical School Ningbo University \\ \hline
    The Second Affiliated Hospital of Jiaxing University \\ \hline
    Department of Rheumatology Huzhou Hospital of Traditional Chinese Medicine Affiliated to Zhejiang Chinese Medical University \\ \hline
    Shaoxing Hospital of Traditional Chinese Medicine, Shaoxing TCM Hospital Affiliated to Zhejiang Chinese Medical University \\ \hline
\end{longtable}

\clearpage 

\subsection*{Acknowledgments} 
\noindent We thank members of the Genomics Platform (Human Phenome Institute, Fudan University) for library preparation experiments and providing the proteomics analysis. We also appreciate the medical background advice provided by Professor Atsushi Ogihara of Waseda University.

\subsection*{Authors' Contributions} 
\noindent LC conceptualized the study, developed the methodology, conducted formal analysis, and led the writing of the manuscript. OD contributed to the development of the methodology, analysis, supervision, and revision of the manuscript. TF, MC, and XZ were involved in data collection, curation, and validation. RC, DL, and RZ provided assistance in manuscript preparation and revision. QJ contributed to conceptualization, supervision, and project administration. XW oversaw the project, providing leadership in conceptualization, supervision, funding acquisition, and manuscript review. All authors have reviewed and approved the final version of the manuscript.

\subsection*{Funding}
\noindent This research was supported by the Zhejiang Traditional Chinese Medical Modernization Special Project (grant number 2020ZX008). 

\subsection*{Data Availability}
\noindent The mass spectrometry proteomics data generated in this study have been deposited in the Integrated Proteome Resources (iProX) database under the dataset identifier IPX0007978000. The data are publicly accessible through the iProX website (https://www.iprox.cn/page/). Additional datasets that support the findings of this study are available from the corresponding author upon reasonable request.

\section*{Declarations}

\subsection*{Ethics approval and consent to participate}
\noindent The study was approved by the ethics committee of the Second Affiliated 58 Hospital of Zhejiang Chinese Medical University (approval NO.2020-KL-002-59 IH01). Informed consent was obtained from all participants.

\subsection*{Consent for publication}
\noindent Not applicable. Our manuscript does not contain any individual person’s data.

\subsection*{Competing Interests}
\noindent The authors declare no competing interests.

\newpage{}




\end{document}